\documentclass[journal,10pt]{IEEEtran} 
\usepackage[figurename=Fig.]{caption}
\usepackage{tikz}
\usetikzlibrary{calc, positioning, shapes.geometric, decorations.pathreplacing}
\usepackage{array}
\usepackage{verbatim}
\usetikzlibrary{calc, shapes, arrows, positioning}
\usepackage{authblk}
\usepackage{blindtext}
\usepackage{ifpdf}
\usepackage{graphicx}
\usepackage{tabulary}
\usepackage{amsmath,amsthm,amssymb}
\usepackage{amsmath}
\usepackage{kantlipsum}
\allowdisplaybreaks
\usepackage{cleveref}
\usepackage{lipsum}%
\usepackage{optidef}
\usepackage[english]{babel} 
\theoremstyle{plain}

\crefname{theorem}{Theorem}{theorem}
\crefname{lemma}{Lemma}{Lemmas}

\usepackage{cite}
\usepackage{dsfont}
\usepackage[justification=centering]{caption}
\usepackage{color}
\usepackage{algorithm}
\usepackage{algorithmicx, algpseudocode}
\usepackage{etoolbox}
\usepackage{subcaption}
\usepackage{balance}
\usepackage{empheq,etoolbox}
\usepackage[utf8]{inputenc}
\usepackage{multirow}
\usepackage{float}

\usepackage{amssymb}
\usepackage{pifont}
\usepackage[hmargin=1.27cm,top=1.3cm,bottom=1.3cm]{geometry}

\usepackage{physics}

\usepackage{babel}
\floatstyle{plaintop}
\restylefloat{table}
\patchcmd{\subequations}
\usepackage{lipsum} 
\allowdisplaybreaks

\tikzset{brace/.style={decorate, decoration={brace}},
 brace mirrored/.style={decorate, decoration={brace,mirror}},
}

\newcounter{brace}
\newcounter{arrow}
\begin{document}
 \captionsetup[figure]{name={Fig.},labelsep=period}

\title{Energy-Efficient Design for Downlink Pinching-Antenna Systems with QoS Guarantee}

\bstctlcite{IEEEexample:BSTcontrol}
\author{Ming Zeng, Ji Wang, Gui Zhou, Fang Fang and Xianbin Wang \textit{Fellow, IEEE}
    \thanks{M. Zeng is with Laval University, Quebec City, Canada (email: ming.zeng@gel.ulaval.ca).}

    
    \thanks{J. Wang is with Central China Normal University, Wuhan, China (e-mail: jiwang@ccnu.edu.cn).}

    \thanks{G. Zhou is with Friedrich-Alexander Universität of Erlangen-Nüremberg, Erlangen, Germany (e-mail: 	gui.zhou@fau.de).}

    \thanks{F. Fang is with Western University, London, Canada (e-mail: fang.fang@uwo.ca).}

    \thanks{X. Wang is with Western University, London, Canada (e-mail: xianbin.wang@uwo.ca).}






    }
\maketitle

\begin{abstract}
Pinching antennas have recently garnered significant attention due to their ability to dynamically reconfigure wireless propagation environments. Despite notable advancements in this area, the exploration of energy efficiency (EE) maximization in pinching-antenna systems remains relatively underdeveloped. In this paper, we address the EE maximization problem in a downlink time-division multiple access (TDMA)-based multi-user system employing one waveguide and multiple pinching antennas, where each user is subject to a minimum rate constraint to ensure quality-of-service. 
The formulated optimization problem jointly considers transmit power and time allocations as well as the positioning of pinching antennas, resulting in a non-convex problem. To tackle this challenge, we first obtain the optimal positions of the pinching antennas. Based on this, we establish a feasibility condition for the system. Subsequently, the joint power and time allocation problem is decomposed into two subproblems, which are solved iteratively until convergence.
Specifically, the power allocation subproblem is addressed through an iterative approach, where a semi-analytical solution is obtained in each iteration. Likewise, a semi-analytical solution is derived for the time allocation subproblem. Numerical simulations demonstrate that the proposed pinching-antenna-based strategy significantly outperforms both conventional fixed-antenna systems and other benchmark pinching-antenna schemes in terms of EE.

\end{abstract}

\IEEEpeerreviewmaketitle

\section{Introduction}

Pinching-antenna technology has recently emerged as a promising solution for establishing line-of-sight (LoS) communication links in high-frequency wireless systems, which are particularly susceptible to LoS blockages due to severe signal attenuation and limited diffraction capabilities \cite{Atsushi_22,  liu2025pinching, yang2025, Hao_Network22}. In a typical pinching-antenna system, the user signals are first fed to an dielectric waveguide, upon which a plastic pinch can be applied to radiate signal, forming an antenna. By dynamically adjusting the pinch location, the wireless propagation environment can be reconfigured in real time, offering substantial flexibility. Prior studies have demonstrated the advantages of pinching-antenna-assisted systems over traditional fixed-antenna architectures across a variety of deployment scenarios \cite{ Xiao25, ding2024,  xie2025, wang2024, hu2025, Li25,  Zeng_COMML25, fu2025, Tegos_2025, zeng2025EE}.
 
Same as conventional systems, resource allocation plays a critical role in unlocking the full potential of pinching-antenna-enabled communications. In addition to conventional wireless resources such as time slots, subcarriers, and transmission power, pinching-antenna systems introduce a new degree of freedom of system design: the pinching antenna position itself. Optimizing this spatial configuration is often tightly coupled with other resource allocation variables, thereby necessitating sophisticated joint optimization approaches. The optimal resource allocation strategy depends on the system objective under consideration. For instance, sum-rate maximization has been widely investigated for both downlink systems using time-division multiple access (TDMA) or non-orthogonal multiple access (NOMA) \cite{ding2024, xie2025, wang2024, hu2025} and uplink systems employing NOMA \cite{Li25, Zeng_COMML25}. Other studies have explored objectives such as total power minimization \cite{fu2025} and minimum rate maximization \cite{Tegos_2025}.

Despite this growing body of work, the problem of energy efficiency (EE) maximization in pinching-antenna systems remains largely unexplored. To the best of our knowledge, the only existing study addressing EE maximization is \cite{zeng2025EE}, which considers a NOMA-based uplink system with a single pinching antenna. In contrast, this paper investigates EE maximization in a downlink system equipped with multiple pinching antennas. To ensure user quality-of-service (QoS), we impose a minimum rate requirement for each user.

We formulate a non-convex optimization problem that jointly considers transmit power allocation, time scheduling, and the spatial configuration of pinching antennas. To address this challenge, we first obtain the optimal antenna positions and establish a feasibility condition for the system. Based on this foundation, the joint power and time allocation problem is decomposed into two subproblems, each solved iteratively until convergence. Specifically, we derive semi-analytical solutions for both power and time allocation subproblems. Numerical results confirm that the proposed approach significantly improves the EE compared to both conventional fixed-antenna systems and state-of-the-art pinching-antenna benchmarks.

\section{System Model and Problem Formulation}
\subsection{System Model}

As illustrated in Fig. 1, we consider a downlink wireless communication scenario, in which a base station (BS) transmits data to multiple users. The BS is equipped with multiple pinching antennas, whose positions can be dynamically adjusted along a single dielectric waveguide. Let $K$ denote the number of users and $N$ the number of pinching antennas. 
Although the system involves multiple antennas and users, conventional multiple-input multiple-output (MIMO) techniques are not applicable due to the constraint that all pinching antennas operate on a shared waveguide, precluding spatial multiplexing. As such, an alternative multiple access scheme is required. Prior works \cite{ding2024, xie2025} demonstrate that when NOMA is employed, the resulting channel gains involve summations of complex numbers, significantly complicating the optimization of antenna placement. 
To circumvent this issue, we adopt TDMA, wherein each user is served in a dedicated time slot. Moreover, to fully exploit the reconfigurability of the pinching antennas, we assume that their positions can be independently optimized for each user, as suggested in \cite{ding2024, xie2025}.

\begin{figure}[ht!]
\centering
\includegraphics[width=1\linewidth]{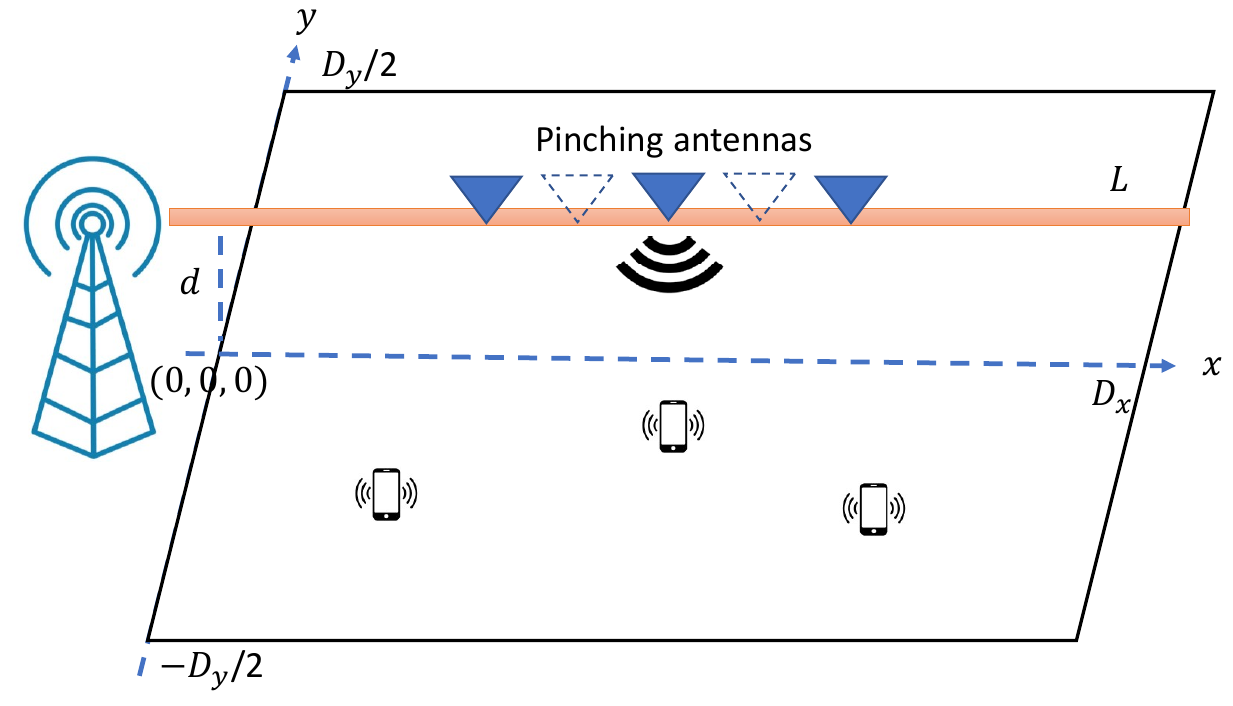}
\caption{A downlink pinching-antenna system.} 
\label{fig:Low_2}
\end{figure}

To clearly describe the system geometry, a three-dimensional Cartesian coordinate system is adopted. The BS is positioned at the origin, i.e., at coordinates $ (0,0,0) $ m. The dielectric waveguide, along which the pinching antennas are deployed, is aligned with the $x$-axis, elevated to a height of $d$ meters above the ground plane, and has a total length of $L$ meters. Users are randomly distributed within a rectangular area on the $xy$-plane, with dimensions $D_x$ and $D_y$. The location of the $k$th user is denoted by $\Phi_{k}=(x_k, y_k, 0) $ m, where $x_k \in [0, D_x]$ and $y_k \in [-\frac{D_y}{2}, \frac{D_y}{2}]$. The position of the 
$n$th pinching antenna, configured specifically for the 
$k$th user, is denoted by $\Phi_{n, k}^{\text{Pin}}=(x_{n, k}^{\text{Pin}}, 0, d) $ m, with $x_{n, k}^{\text{Pin}} \in [0, L] $ m.


Let $s_k$ denote the transmitted symbol intended for user 
$k$, normalized such that $\mathbb{E}{[s_ks_k^H]}=1 $. Since all $N$ pinching antennas are applied to the same waveguide, they effectively form a linear array. As a result, the corresponding signal vector for user $k$ can be expressed as follows, based on the array response model described in \cite{ding2024, xie2025}:
\begin{equation} \label{user signal}
    \mathbf{s}_k=\sqrt{\frac{P_k}{N}} [e^{-j \theta_{1, k}} ~\cdot \cdot \cdot~ e^{-j \theta_{N, k}}]^T s_k, 
\end{equation}
where $P_k$ denotes the transmit power for user $k$, satisfying $P_k \leq P_{\max}$, with $P_{\max} $ being the maximum power constraint. Following the approach in \cite{ding2024, xie2025}, it is assumed that the total transmit power $P_k$ is equally distributed across the $N$ pinching antennas. The phase shift experienced at the $n$th pinching antenna for user $k$ is given by
\begin{equation}
    \theta_{n, k} = 2 \pi \times \frac{ \abs{ \Phi_{n, k}^{\text{Pin}}- \Phi_{0}^{\text{Pin}}} }{ \lambda_g }, 
\end{equation}
where $\Phi_{0}^{\text{Pin}}=(0, 0, d) $ m represents the location of the waveguide’s feed point, while $\lambda_g =\frac{ \lambda}{n_{\text{eff}}} $ is the guided wavelength within the dielectric waveguide. Here, $\lambda$ represents the free-space carrier wavelength, and $n_{\text{eff}} $ is the effective refractive index of the dielectric material. The notation $\abs{ ~\cdot~} $ denotes the Euclidean distance.


We now examine the user channels, which are modeled using the spherical wave propagation model as in \cite{ding2024}. For user $k$, the corresponding channel vector $\mathbf{h}_k \in \mathbb{C}^{N \times 1}$ can be expressed as
\begin{equation} \label{user channel}
    \mathbf{h}_k= \left[ \frac{\eta^{\frac{1}{2}} e^{-j \frac{2\pi}{\lambda} \abs{\Phi_{1, k}^{\text{Pin}}- \Phi_k} } }{\abs{\Phi_{1, k}^{\text{Pin}}- \Phi_k}} \cdot \cdot \cdot 
    \frac{\eta^{\frac{1}{2}} e^{-j \frac{2\pi}{\lambda} \abs{\Phi_{N, k}^{\text{Pin}}- \Phi_k}}  }{\abs{\Phi_{N, k}^{\text{Pin}}- \Phi_k}}
    \right]^T, 
\end{equation}
where $\eta=\frac{c^2}{16 \pi^2 f_c^2}$, $c$ is the speed of light, and $f_c$ is the carrier frequency. 

By combining the user signal expression and the channel model in \eqref{user channel}, the received signal at user 
$k$ can be written as: 
\begin{subequations}
    \begin{align}
        y_k&= \mathbf{h}_k^T \mathbf{s}_k + w_k \\
        &=\left( \sum_{n=1}^N \frac{\eta^{\frac{1}{2}} e^{-j \frac{2\pi}{\lambda} \abs{\Phi_{n, k}^{\text{Pin}}- \Phi_k} }  }{\abs{\Phi_{n, k}^{\text{Pin}}- \Phi_k}} e^{-j \theta_{n,k}} \right) \sqrt{\frac{P_k}{N}} s_k+ w_k, 
    \end{align}
\end{subequations} 
where $w_k$ denotes the additive Gaussian white noise (AWGN) at user $k$, modeled as a complex Gaussian random variable with zero mean and variance $\sigma_k^2$. 

In prior works \cite{ding2024, xie2025}, equal time allocation among users is assumed for simplicity. In contrast, this paper considers a more flexible model, where the time fraction allocated to each user is treated as a variable. Let $\tau_k$
denote the time allocation for user $k$, subject to the constraint $\sum_{k=1}^K \tau_k \leq 1$. For notational simplicity, we define $h_k$ as the effective channel gain normalized by the noise power at user $k$, given by
\begin{equation}
    h_k= \frac{1}{N \sigma_k^2}  \abs{\sum_{n=1}^N \frac{\eta^{\frac{1}{2}} e^{-j \frac{2\pi}{\lambda} \abs{\Phi_{n, k}^{\text{Pin}}- \Phi_k} }  }{\abs{\Phi_{n, k}^{\text{Pin}}- \Phi_k}} e^{-j \theta_{n,k}}}^2. 
\end{equation}

Accordingly, the achievable rate of user $k$ is given by
\begin{equation}
    R_k=\tau_k \times \log_2(1+ P_k h_k). 
\end{equation}

\subsection{Problem Formulation}
As shown in \cite{xie2025}, for the sum-rate maximization problem, the optimal strategy is for each user to transmit at maximum power. However, such an approach can result in excessive power consumption, which is undesirable in energy-constrained systems. To strike a balance between spectral efficiency and energy consumption, this paper focuses on maximizing the system's EE, defined as the ratio of the sum rate to the total power consumption, which includes both the fixed circuit power and the dynamic transmit power \cite{ Zhang_TVT17, Zeng_TVT19}. 
To guarantee QoS for all users, we assume that each user must satisfy a minimum rate requirement. The resulting EE maximization problem entails a joint optimization over the users' transmit powers, time allocations, and the spatial positions of the pinching antennas. The problem is formulated as follows:
\begin{subequations} \label{Problem_formulation}
   \begin{align}
    \max_{x_{n, k}^{\text{Pin}}, P_k, \tau_k}~ &  \frac{\sum_{k=1}^K \tau_k \times \log_2(1+ P_k h_k) }{P_f+ \sum_{k=1}^K P_k} \\
    \text{s.t.}~& x_{n, k}^{\text{Pin}} \in [0, L], \forall n, \forall k\\
    & |x_{n, k}^{\text{Pin}} - x_{n-1, k}^{\text{Pin}}| \geq \Delta_{\min}, \forall n, \forall k\\
    & P_k \in [0, P_{\max}] , \forall k \\
    & \tau_k \geq 0, \forall k \\
    & \sum_{k=1}^K \tau_k \leq 1, \\
    & \tau_k \times \log_2(1+ P_k h_k) \geq R_k^{\min}, \forall k. 
\end{align}  
\end{subequations}
Here, $\Delta_{\min}$ is the minimum distance for the pinching antennas to avoid mutual coupling, while $P_f$ represents the fixed circuit power consumption. Constraint (\ref{Problem_formulation}b) and (\ref{Problem_formulation}c) ensure that each pinching antenna is positioned within the valid range of the waveguide, but adjacent antennas are placed beyond the minimum distance. Constraint (\ref{Problem_formulation}d) enforces the per-user transmit power limits. Constraints (\ref{Problem_formulation}e) and (\ref{Problem_formulation}f) regulate the allocation of time resources among users, while (\ref{Problem_formulation}g) ensures that each user's minimum rate requirement is satisfied.

\section{Proposed Solution}
Problem \eqref{Problem_formulation} is inherently non-convex, primarily due to the non-convex nature of both the objective function and constraint (\ref{Problem_formulation}g). Furthermore, the optimization variables---namely, the transmit power $P_k$, time allocation $\tau_k$, and pinching-antenna locations $x_{n, k}^{\text{Pin}}$---are intricately coupled in both the objective and constraint (\ref{Problem_formulation}g), which significantly complicates the problem. 
To address this challenge, we begin by decoupling the optimization by first focusing on the placement of the pinching antennas.

\subsection{Optimization of Pinching Antennas' Locations}
Since TDMA is employed, there is no inter-user interference, allowing the optimization for each user to be considered independently. Furthermore, it is assumed that the positions of the pinching antennas can be reconfigured individually for each user. Under this assumption, to maximize the effective channel gain for a given user, the corresponding pinching antennas should be positioned as close as possible to the user while ensuring constructive signal combining.
Specifically, this constructive condition can be satisfied by aligning the phases such that
\begin{subequations}
    \begin{align}
        &\frac{2 \pi}{\lambda} \abs{\Phi_{n, k}^{\text{Pin}} -\Phi_k} + \theta_{n, k} \\
        &= \frac{2 \pi}{\lambda} \sqrt{ ( x_{n, k}^{\text{Pin}}- x_k)^2 + y_k^2+d^2 } +\frac{2 \pi}{\lambda_g}  x_{n, k}^{\text{Pin}} \\
        &= 2m \pi,
    \end{align}
\end{subequations}
where $m \in \mathbb{Z}$ is an integer \cite{ding2024}. Based on this condition, the channel gain for user $k$ can be approximated as \cite{ding2024, xie2025}
\begin{subequations}
\begin{align}
  h_k& = \frac{1}{N \sigma_k^2} \times \left( \sum_{n=1}^N \frac{ \eta^{\frac{1}{2}}}{\abs{\Phi_k- \Phi_{n, k}^{\text{Pin}} }} \right)^2   \\
  &\approx \frac{\eta N }{  \abs{\Phi_k- \Phi_{n, k}^{\text{Pin}} }^2 \sigma_k^2}. 
\end{align}
\end{subequations} 

Once the optimal positions of the pinching antennas are determined based on the channel gain maximization criteria, the original problem in \eqref{Problem_formulation} can be simplified by treating the antenna locations as fixed parameters. Consequently, the problem reduces to a joint optimization of transmit power and time allocation as follows:
\begin{subequations} \label{Problem_formulation_2}
   \begin{align}
    \max_{ P_k, \tau_k}~ &  \frac{\sum_{k=1}^K \tau_k \times \log_2(1+ P_k h_k) }{P_f+ \sum_{k=1}^K P_k} \\
    \text{s.t.}~& P_k \in [0, P_{\max}] , \forall k \\
    & \tau_k \geq 0, \forall k \\
    & \sum_{k=1}^K \tau_k \leq 1, \\
    & \tau_k \times \log_2(1+ P_k h_k) \geq R_k^{\min}, \forall k. 
\end{align} 
\end{subequations}


Despite the simplification, problem \eqref{Problem_formulation_2} remains non-convex due to the non-convexity of both the objective function and constraint (\ref{Problem_formulation_2}e). To tackle this, we employ the block coordinate descent (BCD) method, which decomposes the original problem into two alternating subproblems: one for power allocation and the other for time allocation. Prior to solving these subproblems, we first assess the feasibility of the reformulated optimization problem.

\subsection{Feasibility of the Considered Problem}
Given that the problem is subject to both transmit power constraints and minimum rate requirements, infeasibility may arise. To verify feasibility, a necessary condition is that each user should have enough time to achieve its minimum rate when transmitting at maximum power, i.e., $P_k=P_{\max}, \forall k$. The required minimum time for user $k$ is given by 
\begin{equation}
    \tau_k^{\min}  = \frac{R_k^{\min} }{\log_2(1+ P_{\max} h_k)  }.
\end{equation}

The problem is feasible if and only if the following condition is satisfied:
\begin{equation}
    \sum_{k=1}^K \tau_k^{\min}  \leq 1. 
\end{equation}

For the subsequent joint optimization of power and time allocation, we proceed under the assumption that the problem is feasible.

\subsection{Power Optimization under Given Time Allocation}
Given a fixed time allocation, the power optimization problem can be reformulated as  
\begin{subequations} \label{Problem_formulation_3}
   \begin{align}
    \max_{ P_k}~ &  \frac{\sum_{k=1}^K \tau_k \times \log_2(1+ P_k h_k) }{P_f+ \sum_{k=1}^K P_k} \\
    \text{s.t.}~& P_k \in [0, P_{\max}] , \forall k \\
    & \tau_k \times \log_2(1+ P_k h_k) \geq R_k^{\min}, \forall k. 
\end{align} 
\end{subequations}

Since the numerator of the objective function is concave and the denominator is affine, the problem constitutes a quasi-concave fractional program. Additionally, all constraints are affine. Therefore, the problem can be efficiently solved using the Dinkelbach algorithm \cite{Dinkelbach}. This method transforms the original fractional problem into a sequence of parameterized subtractive-form subproblems, each of which can be expressed as follows:
\begin{subequations} \label{Problem_formulation_32}
   \begin{align}
    \max_{ P_k}~ &  {\sum_{k=1}^K \tau_k \times \log_2(1+ P_k h_k) }- \beta^{(i-1)} \bigg( {P_f+ \sum_{k=1}^K P_k} \bigg) \\
    \text{s.t.}~& P_k \in \bigg[\frac{2^{ \frac{R_k^{\min}}{\tau_k}}-1 }{h_k}, P_{\max} \bigg] , \forall k,  
\end{align} 
\end{subequations}
where constraint (\ref{Problem_formulation_32}b) is the combination of constraints (\ref{Problem_formulation_3}b) and (\ref{Problem_formulation_3}c). Additionally, $\beta^{(i-1)}$ denotes EE value obtained at the $(i-1)$th iteration. The initialization of $\beta$ is set as 
\begin{equation}
    \beta^{(0)}=\frac{\sum_{k=1}^K \tau_k \times \log_2(1+ P_{\max} h_k) }{P_f+ \sum_{k=1}^K P_{\max}},
\end{equation}
and the update of $\beta^{(i)} $ at iteration $i$ is given by
\begin{equation}
    \beta^{(i)}=\frac{\sum_{k=1}^K \tau_k \times \log_2(1+ P_{k}^{(i)} h_k) }{P_f+ \sum_{k=1}^K P_{k}^{(i)}}.  
\end{equation} 
In the above expression, $P_{k}^{(i)}$ denotes the updated transmit power obtained by solving problem \eqref{Problem_formulation_32} at the $i$th iteration. As established in \cite{Dinkelbach}, the parameter $\beta^{(i)} $
is guaranteed to be non-decreasing with each iteration. The iterative procedure terminates once the increment $\beta^{(i)}-\beta^{(i-1)}$ falls below a predefined convergence threshold, typically set to $10^{-6}$. At convergence, the final value 
$\beta^{(i)} $ corresponds to the globally optimal EE for problem \eqref{Problem_formulation_3}.


The challenge now lies in solving problem \eqref{Problem_formulation_32}. Given the parameter 
$\beta^{(i-1)}$, this sum problem can be decomposed into 
$K$ independent parallel subproblems. The optimal transmit power for each user is then determined as follows:
\begin{equation}
    P_k^{\text{Opt}}=
    \begin{cases}
    P_k^{\text{Der}}, \text{if}~ P_k^{\text{Der}} \in \left[\frac{2^{ \frac{R_k^{\min}}{\tau_k}}-1 }{h_k}, P_{\max} \right]\\
    \frac{2^{ \frac{R_k^{\min}}{\tau_k}}-1 }{h_k}, \text{if}~ P_k^{\text{Der}} <\frac{2^{ \frac{R_k^{\min}}{\tau_k}}-1 }{h_k}\\
    P_{\max}, \text{if}~ P_k^{\text{Der}} >P_{\max},
    \end{cases}
\end{equation}
where $ P_k^{\text{Der}}= \frac{\tau_k}{\beta^{(i-1)} \ln{2} }- \frac{1}{h_k}$ is the stationary point of problem \eqref{Problem_formulation_32} with respect to $P_k$.  

\subsection{Time Optimization under Given Power Allocation}
Given a fixed power allocation, the time allocation problem can be formulated as:
\begin{subequations} \label{Problem_formulation_4}
   \begin{align}
    \max_{  \tau_k}~ &  {\sum_{k=1}^K \tau_k \times \log_2(1+ P_k h_k) } \\
    \text{s.t.}~
    & \sum_{k=1}^K \tau_k \leq 1, \\
    & \tau_k \times \log_2(1+ P_k h_k) \geq R_k^{\min}, \forall k. 
\end{align} 
\end{subequations}

Note that constraint (\ref{Problem_formulation}e) is omitted here since it is inherently guaranteed by constraint (\ref{Problem_formulation_4}c). Observing the linear objective and constraints, it follows that the optimal time allocation strategy is to assign each user---except the one with the largest achievable rate under unit time, defined as $ \log_2(1+ P_k h_k)$---the minimum time required to meet its rate constraint with equality. The remaining time is allocated to the user with the highest value of $ \log_2(1+ P_k h_k)$, thereby maximizing the overall objective.

Without loss of generality, denote user $l$ as the user with the largest value of $ \log_2(1+ P_k h_k), \forall k$. Accordingly, the optimal time allocation satisfies
\begin{equation}
    \tau_k^{\text{Opt}}=
    \begin{cases}
    \frac{R_k^{\min}}{ \log_2(1+ P_k h_k) }, \text{if}~ k \neq l \\
    1- \sum_{i=1, i \neq l}^K \tau_i^{\text{Opt}}, \text{else}.
    \end{cases}
\end{equation}

\subsection{Iterative Update until Convergence}
The power and time allocation subproblems are solved iteratively until convergence. Given that each subproblem is optimally solved at every iteration, the EE metric is guaranteed to be non-decreasing after each update. Consequently, the iterative algorithm converges to a stationary point.

\begin{figure}[ht!]
\centering
\includegraphics[width=1\linewidth]{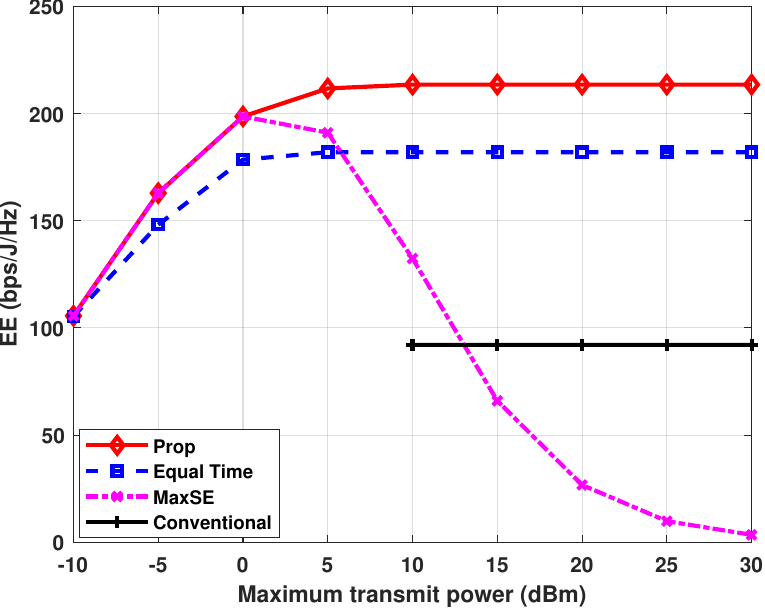}
\caption{EE versus the maximum transmit power constraint at the users.} 
\label{Power}
\end{figure}

\begin{figure}[ht!]
\centering
\includegraphics[width=1\linewidth]{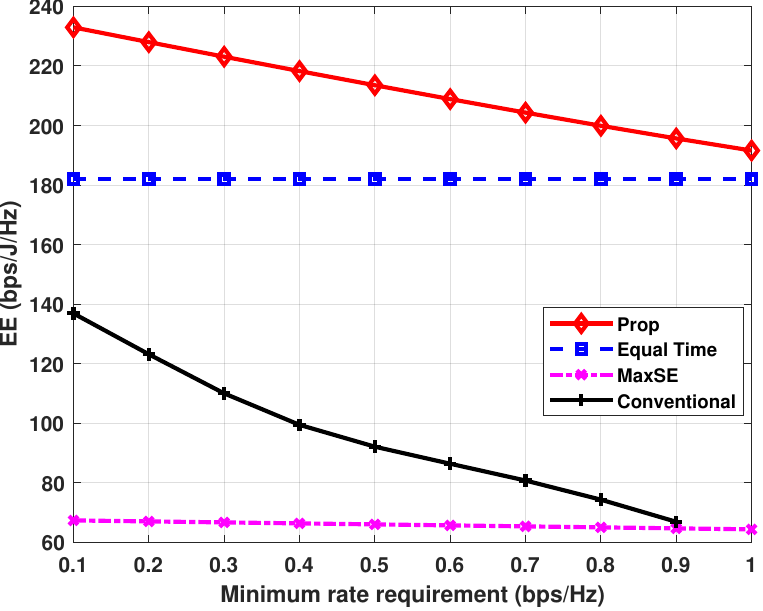}
\caption{EE versus the minimum rate requirement at the users.} 
\label{QoS}
\end{figure}

\section{Numerical Results}
Simulation results are presented to demonstrate the effectiveness of the proposed scheme (denoted as ``Prop''). The default simulation parameters are set as follows: the maximum transmit power constraint and fixed circuit power consumption are both 15 dBm, and the carrier frequency is 28 GHz. The system serves $K=5$ users, whose locations are randomly distributed within a service area of dimension $D_x=60$ m and $D_y=20$. Each user’s minimum rate requirement is $ R_k^{\min}=0.5$ bps/Hz, $\forall k$, and the noise variance at each user is $\sigma_k^2=-90 $ dBm, $\forall k$. Consistent with \cite{ding2024, xie2025}, the length of the dielectric waveguide is set equal to the horizontal span of the service region, i.e., $L=D_x$. All simulation results are averaged over 
$10^4$ independent Monte Carlo realizations.


For comparison, three benchmark schemes are considered: 1) Conventional fixed-antenna configuration (denoted as ``Conventional''), where a uniform linear array of $N$ antenna is placed along the x-axis, starting from the position (0,0,d) m, with half-wavelength spacing.
2) Sum rate maximization strategy (denoted as ``MaxSE'') from \cite{xie2025}, in which each user transmits at maximum power; and 3) Equal time allocation scheme (denoted as ``Equal Time'') based on \cite{ding2024, xie2025}, where equal time resources are allocated to all users.


Figure \ref{Power} illustrates the EE as a function of the maximum transmit power constraint $P_{\max}$ for the evaluated schemes. Across the entire range of $P_{\max}$ values considered, all schemes employing pinching antennas consistently satisfy the QoS requirements, whereas the conventional fixed-antenna configuration fails to meet these requirements when $P_{\max} <10$ dBm. This highlights the clear advantage of the pinching-antenna system over the conventional fixed setup. As $P_{\max}$ increases, the EE performance of both the proposed scheme (``Prop'') and the equal time allocation scheme (``Equal Time'') initially improves before saturating. Conversely, the EE of the sum rate maximization scheme (``MaxSE'') initially increases but subsequently declines with further increases in $P_{\max}$. This behavior demonstrates that maximizing EE offers a more effective trade-off between throughput and power consumption compared to sum rate maximization. For any given $P_{\max}$, the ``Prop'' scheme outperforms ``Equal Time'' in terms of EE, with the performance gap attributable to the optimized time allocation among users.


Figure \ref{QoS} depicts the variation of EE with respect to the minimum rate requirement $R^{\min}$. As expected, increasing $R^{\min}$ generally results in a decrease or stabilization of EE. This decreasing trend is evident for both the proposed scheme (``Prop'') and the conventional fixed-antenna configuration (“Conventional”), while the EE performance of the other two schemes remains relatively insensitive to changes in $R^{\min}$. Notably, the ``Prop'' scheme consistently achieves the highest EE across the range of minimum rate requirements, further validating its superior effectiveness.

\begin{figure}[ht!]
\centering
\includegraphics[width=1\linewidth]{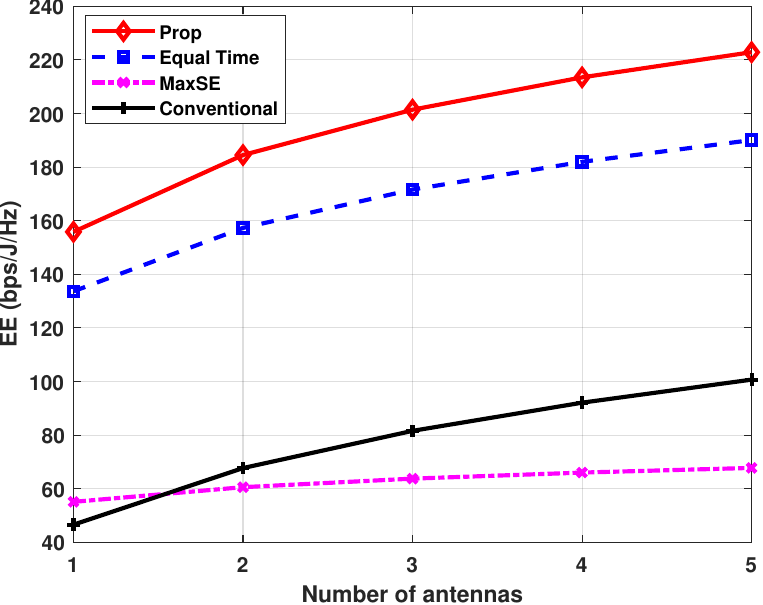}
\caption{EE versus the number of pinching antennas.} 
\label{antenna}
\end{figure}

Finally, Fig. \ref{antenna} illustrates the impact of the number of antennas on the system’s EE. As anticipated, EE improves with an increasing number of antennas for all evaluated schemes. Among them, the ``Prop'' scheme outperforms all others, followed by the ``Equal Time'' scheme. The EE curves for the ``MaxSE'' and ``Conventional'' schemes intersect, with their relative performance ranking depending on the specific antenna count.



\section{Conclusion} 
\label{Sec:Conclusion}
In this paper, we investigated the EE maximization for a downlink TDMA system equipped with multiple pinching antennas, subject to individual user minimum rate constraints. The formulated optimization problem involved the joint design of user transmit power, time allocation, and the spatial positioning of the pinching antennas. Due to the coupled and non-convex nature of the problem, we first obtained the optimal antenna positions and established a corresponding feasibility condition. Based on this, the joint power and time allocation subproblems were addressed iteratively using the BCD method, with semi-analytical solutions developed for each subproblem. Numerical results demonstrated that pinching-antenna-assisted systems more effectively satisfied users’ QoS requirements compared to conventional fixed-antenna-based systems. Furthermore, the proposed scheme consistently outperformed both fixed-antenna baselines and existing pinching-antenna benchmarks in terms of EE.


\bibliographystyle{IEEEtran}
\bibliography{biblio}

\begin{thebibliography}{10}
\providecommand{\url}[1]{#1}
\csname url@samestyle\endcsname
\providecommand{\newblock}{\relax}
\providecommand{\bibinfo}[2]{#2}
\providecommand{\BIBentrySTDinterwordspacing}{\spaceskip=0pt\relax}
\providecommand{\BIBentryALTinterwordstretchfactor}{4}
\providecommand{\BIBentryALTinterwordspacing}{\spaceskip=\fontdimen2\font plus
\BIBentryALTinterwordstretchfactor\fontdimen3\font minus \fontdimen4\font\relax}
\providecommand{\BIBforeignlanguage}[2]{{%
\expandafter\ifx\csname l@#1\endcsname\relax
\typeout{** WARNING: IEEEtran.bst: No hyphenation pattern has been}%
\typeout{** loaded for the language `#1'. Using the pattern for}%
\typeout{** the default language instead.}%
\else
\language=\csname l@#1\endcsname
\fi
#2}}
\providecommand{\BIBdecl}{\relax}
\BIBdecl

\bibitem{Atsushi_22}
A.~Fukuda, H.~Yamatoto, H.~Okazaki, Y.~Suzuki, and K.~Kawai, ``Pinching antenna using a dielectric waveguide as an antenna,'' \emph{Technical Journal}, vol.~23, no.~3, pp. 5--12, 2022.

\bibitem{liu2025pinching}
\BIBentryALTinterwordspacing
Y.~Liu, Z.~Wang, X.~Mu, C.~Ouyang, X.~Xu, and Z.~Ding, ``Pinching-antenna systems ({PASS}): Architecture designs, opportunities, and outlook,'' 2025. [Online]. Available: \url{https://arxiv.org/abs/2501.18409}
\BIBentrySTDinterwordspacing

\bibitem{yang2025}
\BIBentryALTinterwordspacing
Z.~Yang, N.~Wang, Y.~Sun, Z.~Ding, R.~Schober, G.~K. Karagiannidis, V.~W.~S. Wong, and O.~A. Dobre, ``Pinching antennas: Principles, applications and challenges,'' 2025. [Online]. Available: \url{https://arxiv.org/abs/2501.10753}
\BIBentrySTDinterwordspacing

\bibitem{Hao_Network22}
W.~Hao, F.~Zhou, M.~Zeng, O.~A. Dobre, and N.~Al-Dhahir, ``Ultra wideband thz {IRS} communications: Applications, challenges, key techniques, and research opportunities,'' \emph{IEEE Network}, vol.~36, no.~6, pp. 214--220, 2022.

\bibitem{Xiao25}
J.~Xiao, J.~Wang, M.~Zeng, Y.~Liu, and Z.~Ding, ``{OFDM-PASS}: Frequency-selective modeling and analysis for pinching-antenna systems,'' \emph{IEEE Wirel. Commun. Lett.}, 2025, submitted.

\bibitem{ding2024}
Z.~Ding, R.~Schober, and H.~Vincent~Poor, ``Flexible-antenna systems: A pinching-antenna perspective,'' \emph{IEEE Transactions on Communications}, pp. 1--1, 2025.

\bibitem{xie2025}
\BIBentryALTinterwordspacing
X.~Xie, F.~Fang, Z.~Ding, and X.~Wang, ``A low-complexity placement design of pinching-antenna systems,'' 2025. [Online]. Available: \url{https://arxiv.org/abs/2502.14250}
\BIBentrySTDinterwordspacing

\bibitem{wang2024}
K.~Wang, Z.~Ding, and R.~Schober, ``Antenna activation for {NOMA} assisted pinching-antenna systems,'' \emph{IEEE Wirel. Commun. Lett.}, pp. 1--1, 2025.

\bibitem{hu2025}
\BIBentryALTinterwordspacing
S.~Hu, R.~Zhao, Y.~Liao, D.~W.~K. Ng, and J.~Yuan, ``Sum-rate maximization for pinching antenna-assisted {NOMA} systems with multiple dielectric waveguides,'' 2025. [Online]. Available: \url{https://arxiv.org/abs/2503.10060}
\BIBentrySTDinterwordspacing

\bibitem{Li25}
Y.~Li, J.~Wang, M.~Zeng, Y.~Liu, and Z.~Ding, ``Sum rate maximization for wireless powered pinching-antenna systems (pass),'' \emph{IEEE Commun. Lett.}, 2025, submitted.

\bibitem{Zeng_COMML25}
\BIBentryALTinterwordspacing
M.~Zeng, J.~Wang, X.~Li, G.~Wang, O.~A. Dobre, and Z.~Ding, ``Sum rate maximization for {NOMA}-assisted uplink pinching-antenna systems,'' 2025. [Online]. Available: \url{https://arxiv.org/abs/2505.00549}
\BIBentrySTDinterwordspacing

\bibitem{fu2025}
\BIBentryALTinterwordspacing
Y.~Fu, F.~He, Z.~Shi, and H.~Zhang, ``Power minimization for noma-assisted pinching antenna systems with multiple waveguides,'' 2025. [Online]. Available: \url{https://arxiv.org/abs/2503.20336}
\BIBentrySTDinterwordspacing

\bibitem{Tegos_2025}
S.~A. Tegos, P.~D. Diamantoulakis, Z.~Ding, and G.~K. Karagiannidis, ``Minimum data rate maximization for uplink pinching-antenna systems,'' \emph{IEEE Wirel. Commun. Lett.}, p. 1–1, 2025, early access.

\bibitem{zeng2025EE}
\BIBentryALTinterwordspacing
M.~Zeng, X.~Li, J.~Wang, G.~Huang, O.~A. Dobre, and Z.~Ding, ``Energy-efficient resource allocation for {NOMA}-assisted uplink pinching-antenna systems,'' 2025. [Online]. Available: \url{https://arxiv.org/abs/2505.07555}
\BIBentrySTDinterwordspacing

\bibitem{Zhang_TVT17}
Y.~Zhang, H.-M. Wang, T.-X. Zheng, and Q.~Yang, ``Energy-efficient transmission design in non-orthogonal multiple access,'' \emph{IEEE Trans. Veh. Tech.}, vol.~66, no.~3, pp. 2852--2857, Mar. 2017.

\bibitem{Zeng_TVT19}
M.~Zeng, N.-P. Nguyen, O.~A. Dobre, Z.~Ding, and H.~V. Poor, ``Spectral- and energy-efficient resource allocation for multi-carrier uplink {NOMA} systems,'' \emph{IEEE Trans. Veh. Tech.}, vol.~68, no.~9, pp. 9293--9296, Mar. 2019.

\bibitem{Dinkelbach}
W.~Dinkelbach, ``On nonlinear fractional programming,'' \emph{Management Science}, vol.~13, no.~7, pp. 492--498, 1967.

\end{thebibliography}

\end{document}